\documentclass[aps,prl,twocolumn,superscriptaddress]{revtex4-1}
\usepackage{fancyhdr} 
\usepackage{color}
\usepackage{hyperref} 
\usepackage{pdfpages} 
\makeatletter
\AtBeginDocument{\let\LS@rot\@undefined}
\makeatother
\usepackage{float}

\usepackage{verbatim} 
\usepackage{amsmath} 
\usepackage{amsthm} 
\usepackage{amssymb}	
\usepackage{graphicx} 

\let\baraccent=\= 
\renewcommand{\=}[1]{\stackrel{#1}{=}} 

\newcommand{\gae}{\lower 2pt \hbox{$\,
    \buildrel{\scriptstyle >}\over {\scriptstyle \sim}\,$}}
\newcommand{\lae}{\lower 2pt \hbox{$\,
    \buildrel{\scriptstyle <}\over {\scriptstyle \sim}\,$}}

\newcommand{\ket}[1]{\big| #1 \big\rangle} 
\newcommand{\bra}[1]{\big\langle #1 \big|} 

\usepackage{xcolor}
\definecolor{violet}{rgb}{0.58, 0.0, 0.83}

\begin{document}

\title{Signatures of information scrambling in the dynamics of the entanglement spectrum} 
\author{T. Rakovszky}
\affiliation{Department of Physics, T42, Technische Universit\"at M\"unchen,
James-Franck-Stra{\ss}e 1, D-85748 Garching, Germany}

\author{S.~Gopalakrishnan}
\affiliation{Department of Physics and Astronomy, CUNY College of Staten Island,
Staten Island, NY 10314, USA}
\affiliation{Physics Program and Initiative for the Theoretical Sciences,
The Graduate Center, CUNY, New York, NY 10016, USA}

\author{S.~A.~Parameswaran}
\affiliation{Rudolf Peierls Centre for Theoretical Physics,  Clarendon Laboratory, University of Oxford, Oxford OX1 3PU, UK}
\author{F. Pollmann}
\affiliation{Department of Physics, T42, Technische Universit\"at M\"unchen,
James-Franck-Stra{\ss}e 1, D-85748 Garching, Germany}
\affiliation{Munich Center for Quantum Science and Technology (MCQST), Ludwig-Maximilians-Universit{\"a}t M{\"u}nchen, Fakult{\"a}t f{\"u}r Physik, Schellingstr. 4, D-80799 M{\"u}nchen, Germany}

\date{\today}

\begin{abstract}
We examine the time evolution of the entanglement spectrum of a small subsystem of a non-integrable spin chain following a quench from a product state. 
We identify signatures in this entanglement spectrum of the distinct dynamical velocities (related to entanglement and operator spreading) that control thermalization. 
We show that the onset of level repulsion in the entanglement spectrum occurs on different timescales depending on the ``entanglement energy'', and that this dependence reflects the shape of the operator front.  
Level repulsion spreads across the entire entanglement spectrum on timescales close to when the mutual information between individual spins at the ends of the subsystem reaches its maximum; this timescale is much shorter than that for full thermalization of the subsystem. We provide an analytical understanding of this phenomenon and show supporting numerical data for both random unitary circuits and a microscopic Hamiltonian.
\end{abstract}
\maketitle

\noindent\textit{Introduction.---}Quantum quenches, which track the dynamics of an isolated quantum system from a simple initial state (e.g., product state), are of widespread experimental relevance in ultracold atomic systems as well as solid-state systems probed on ultrafast timescales~\cite{bdz, wang2013, polkovnikov_review}. How such systems approach local thermal equilibrium after a quench, i.e., ``thermalization'', is a central theme in many-body physics~\cite{Rigol2008,CalabreseCardy06,ETHreviewRigol16,GogolinReview,Kaufman794}. 
While the late-time behavior of small subsystems is essentially thermal~\cite{Deutsch91,Srednicki94,ETHreviewRigol16}, less is known about early and intermediate times, i.e., how the local density matrix morphs from a product state to a thermal state.  Characterizing the intermediate, locally thermal regime is a key step, both for understanding thermalization and for devising efficient numerical methods to study dynamics~\cite{Leviatan2017}. 
Different aspects of quantum information propagate at distinct and well-separated speeds~\cite{Mezei16, tianci, ho2017}, so a local subsystem has a spectrum of thermalization timescales, and exhibits rich intermediate-time structures.
\begin{figure}[h!]
\begin{center}
\includegraphics[width = \columnwidth]{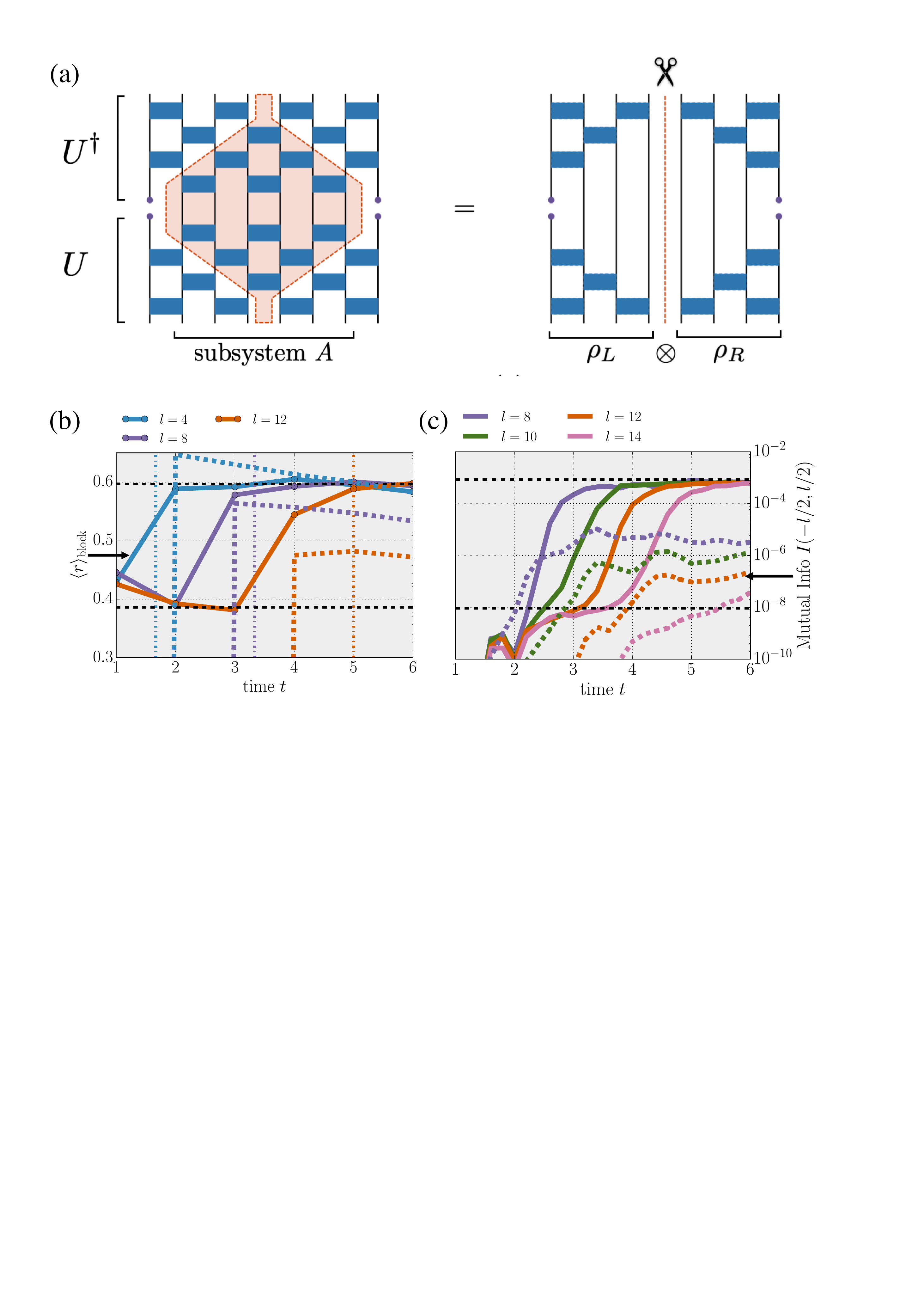}
\caption{(a)~Representation of the reduced density matrix $\rho_{\bar{A}}$ after a few steps of time evolution under a local unitary circuit. Tracing over subsystem $A$ causes unitaries to cancel inside the red shaded region. Consequently, $\rho_{\bar{A}}$ factors as a tensor product, and the entanglement spectrum (ES) decouples into left/right contributions. (b) ES dynamics for a random unitary circuit, showing the linear subsystem-size-dependent crossover from Poisson to RMT level statistics of the ES (solid) and mutual information between spins on either side of subsystem $A$ (dashed); (c) shows similar data for Hamiltonian dynamics of the nonintegrable Ising model~\eqref{eq:Isingham}. In both cases the level statistics are computed for entanglement energies $E\leq 10$.}
\label{fig:mainresults}
\end{center}
\end{figure}

In this work, we explore this intermediate-time regime and tease apart various dynamical timescales. We study the spectrum of the reduced density matrix (RDM) $\rho_A$ of a subsystem $A$ of size $l \gg 1$ that is small compared to system size $L$, at times $t$ that are short compared with the timescale for the full thermalization of $\rho_A$~\cite{Mezei16,JonayNahum}. In this regime, $\rho_A$ is not thermal, but chaos is expected on length-scales smaller than $l$, so one expects some aspects of the dynamics to be universal. Further, since this regime involves weakly entangled states, large-scale simulations using matrix-product states (MPSs) are feasible. 
We argue that the spectral correlations of $\rho_A$ identify the timescales involved in thermalization. These correlations are often quantified 
via the \emph{entanglement spectrum}, i.e. the eigenvalues of the so-called entanglement Hamiltonian $H_\text{ent} \equiv -\ln{\rho_A}$. The entanglement spectrum was originally introduced~\cite{LiHaldane08} as a powerful tool for characterizing ground states; more recently, it has been used in non-equilibrium settings~\cite{ChamonHamma1,ChamonHamma2,Chamon18,Serbyn16,RUCEntSpectra}. We continue in this vein and explore its post-quench dynamics. Following a quench, degrees of freedom near either end of $A$ quickly become entangled with the outside world, but take longer to become entangled with those at the other end of $A$. This is reflected in the spectrum of $\rho_A$: its large eigenvalues (i.e., low-energy part of the entanglement spectrum) correspond to eigenstates that are localized on either end of the system, leading to Poissonian level statistics on short timescales, whereas small eigenvalues of 
$\rho_A$ (i.e., high-energy part of the entanglement spectrum) couple and become essentially random on much shorter times. We examine the crossover between these two sectors of the entanglement spectrum with time, and link it with the spread of operators and entanglement across the subsystem.

\textit{Analytical argument}.--- We begin by providing an intuitive analytical picture of the dynamics of the entanglement spectrum. For clarity, let us focus on random unitary circuit (RUC) dynamics~\cite{Nahum16,Nahum17,RvK17,tianci,RUCEntSpectra}, where the existence of a strict light cone velocity $v_{\text{LC}}$ streamlines the discussion. 
We generate time evolution under RUCs in terms of two-site unitary gates that act on even (odd) bonds at integer (half-integer) time steps as shown in Fig.~\ref{fig:mainresults}(a), so that $v_{\text{LC}} =2$; note that such a representation can also be used to approximate Hamiltonian dynamics to arbitrary accuracy via a Trotter decomposition. Apart from the light cone velocity, the random circuit has two other distinct characteristic scales, the \emph{butterfly velocity} $v_\text{B}$, which sets how fast operators spread in space, and the \emph{entanglement velocity} $v_\text{E}$, related to the speed of entanglement growth when the chain is partitioned into two halves; these satisfy $v_\text{LC} > v_\text{B} > v_\text{E}$~\cite{Mezei16, Nahum17, RvK17}. 

We begin with a pure product state at $t=0$ and evolve it to time $t$ by applying a depth-$t$ RUC. The RDM $\rho_{A}$ of subsystem $A$ is obtained by constructing the density matrix of the whole system, $\rho = |\psi(t) \rangle \langle \psi(t)|$, and then tracing over degrees of freedom outside $A$.
However, since the spectrum of $\rho_{A}$ is identical  (up to zero modes) to that of the RDM of the complement of $A$ (denoted $\rho_{\bar{A}}$) we may instead perform the trace over the degrees of freedom within $A$, corresponding to the circuit on the LHS of Fig.~\ref{fig:mainresults}(a), where the purple dots denote the degrees of freedom in $\bar{A}$.
 After canceling conjugate pairs of gates $U$ and $U^\dagger$, this circuit separates into a tensor product: $\rho_{\bar{A}} = \rho_L \otimes \rho_R$, where $L/R$ denote regions to the left and right of $A$ respectively. 
Therefore the eigenvalues of $\rho_{\bar{A}}$, and hence those of $\rho_A$, take the form $\lambda^\alpha_L \lambda^\beta_R$, and the entanglement spectrum is the sum of the spectra of independent random matrices, leading to Poisson level statistics. After an initial non-universal transient (from the singular entanglement spectrum of each edge in the initial state) we expect such behavior up to time $l/2v_{\text{LC}}$.
  
When $t > l/2v_{\text{LC}}$, this picture changes: canceling pairs of conjugate unitaries no longer partitions the circuit into disjoint pieces, so the RDM no longer factorizes. However, even for $t > l/2v_{\text{LC}}$ the left and right blocks are initially only weakly entangled, since any entanglement between them is produced only by the few gates between the corners of the light-cone region and the bottom/top of the circuit. This idea can be formalized by considering the mutual information between the regions $L$ and $R$ that are left and right of $A$: ${I} (L,R) = S_L + S_R - S_{\bar{A}}$. At short times, when the RDM of $\bar{A}$ factorizes, ${I} (L,R) = 0$; at later times, one can bound  ${I} (L,R) \leq 2 v_{\text{LC}} t - l$. In our two-level circuits, the ``butterfly velocity'' $v_B$ that characterizes the spread of operators, is much slower than $v_{\text{LC}}$, so on times comparable to $v_{\text{LC}}$, the inequality is not saturated. 
On these timescales, only rare low-amplitude operators entangle the two halves of the subsystem, which is thus almost separable.

Suppose we approximate an almost separable density matrix as $\widetilde{\rho} = \rho_L \otimes \rho_R + \Delta \rho$. The perturbation $\Delta \rho$ will induce stronger hybridization among levels that are closer together in energy. Both $\rho_L$ and $\rho_R$ have a density of states $N(\lambda) \propto 1/\lambda$~\cite{RUCEntSpectra}. {States at small $\lambda$, which correspond to high entanglement energies, are tightly packed, and therefore more susceptible to being hybridized by $\Delta \rho$ than the low-energy, well-separated ones}. 
As entanglement grows, the entanglement spectrum shows random matrix behavior at successively lower energies, until at a time  $t\approx l/2v^*$, all the levels show RMT statistics. We conjecture that $v^* =v_B$~\cite{suppmat}.  Note that this time is still lower than the time $t\approx l/2v_E$ at which the entanglement entropy of $A$ saturates to its thermal value, since on general grounds it is expected that $v_E < v_B$~\cite{Mezei16, JonayNahum}. This final stage of thermalization is associated with the gradual evolution of the entanglement density of states to the semicircle distribution characteristic of random matrix theory (RMT). Before this time, although the entanglement spectrum shows RMT statistics, its weighted average (i.e., the von Neumann entropy) has yet to reach its asymptotic thermal volume law value (that is $\propto l$).

These ideas apply immediately to Floquet circuits~\cite{ChanDeLuca1,ChanDeLuca2,ChanDeLuca3} (unitary circuits that are random in space but are time-periodic), as these  also have a strict light cone. Their adaptation to more general Floquet problems or to Hamiltonian dynamics is straightforward~\cite{suppmat}:  such models are equivalent to time evolution with a strictly local circuit up to small, longer-range terms~\cite{OsborneApprox}, which exclusively affect high entanglement energies. The key distinction is that formally there is no strict light cone speed bounding the time at which level statistics are affected; however, in practice for any given accuracy we may choose   $v_{\text{LC}} = v_\text{LR}$, a Lieb-Robinson speed, as such a bound.
The early-time behavior is significantly modified when both the dynamics and the initial state are translation invariant. In this case the spectra of the left and right edges are identical, which results in exact degeneracies, rather than Poisson statistics, at early times~\cite{suppmat}.

\textit{Numerical Simulations}.--- Our dynamical regime of interest  consists of not too large subsystems at intermediate times. Since the relevant time evolution only generates modest entanglement, it is feasible to simulate it efficiently using matrix-product state (MPS) techniques. We simulate dynamics using the time-evolving block decimation (TEBD) algorithm~\cite{VidalTEBD} on systems of size $L=60$, and compute the entanglement spectra of subsystems ranging in size from $l=4$ to $l=14$. Note that the limiting factor in going to larger $l$ lies in the fact that, unlike in typical applications of MPS technology, we are interested in the entire entanglement spectrum rather than its low-energy sector.
We have verified that $L$ is sufficiently large that it does not produce any significant finite-size effects; all finite-size scaling is controlled by $l$.
For RUCs, we draw two-site unitary gates at random from the Haar measure, and average results over 100 realizations. Here, the time $t$ counts the number of full time steps in which each even and each odd bond is acted upon exactly once by a 2-site gate. For the Hamiltonian case, we study the  Ising model in a tilted magnetic field, 
\begin{equation}\label{eq:Isingham}
H = \sum\nolimits_i J_i \sigma^x_i\sigma^x_{i+1} +  h^z_i \sigma^z_i + h^x_i \sigma^x_i,
\end{equation}
where $\sigma^\mu_i$ with $\mu=x,y,z$ are Pauli matrices at lattice site $i$. We chose $J_i = J = 1$ to be uniform and measure time in units of $1/J$. To avoid dealing with subtleties of thermalization within symmetry sectors, and to make the two edges of the block inequivalent, we add weak on-site disorder, taking $h^{z,x} \in [\overline{h^{z,x}}-\frac{W}{2},\overline{h^{z,x}}+\frac{W}{2}]$ and choose their averages to be $\overline{h^{z}} = 0.9045$, $\overline{h^{z}} = 0.709$ and the disorder strength $W=0.05$; the tilted-field Ising chain is known to be ergodic for this choice~\cite{HyungwonHuse}. We average the results over $50$ disorder realizations. For both models, we begin with an initial N\'eel state $\ket{\Psi(t=0)} = \ket{\uparrow\downarrow\uparrow\downarrow\ldots}$.

\begin{figure}[t]
\begin{center}
\includegraphics[width = .9\columnwidth]{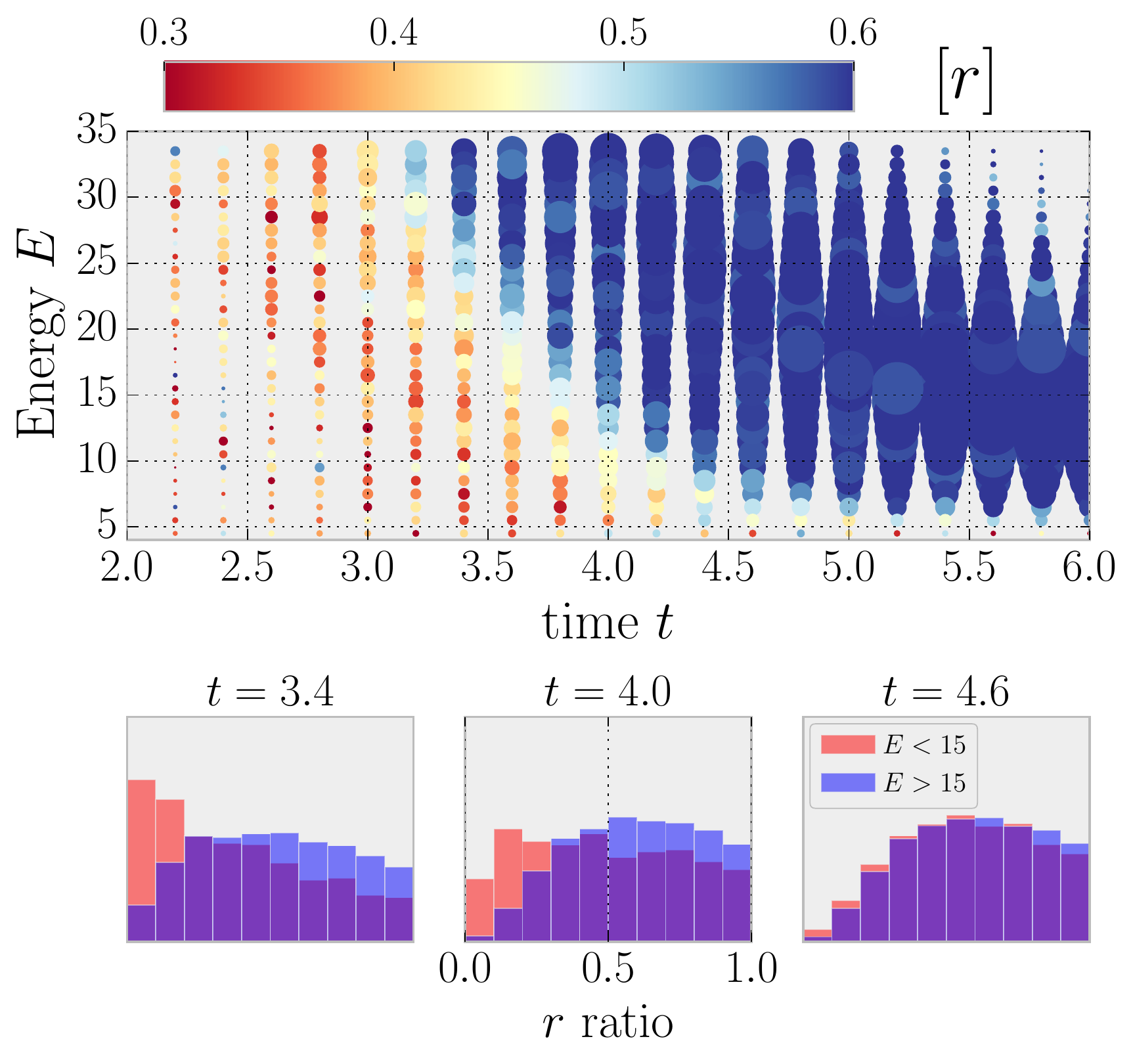}
\caption{Time evolution of energy-resolved ES level statistics for the non-integrable Ising model shows successively lower-energy states crossing over from Poisson to RMT behavior. The color of each dot corresponds to the $r$-ratio and the size to the number of states in the energy window of size $\Delta E = 1$. Bottom:  histograms of low/high energy parts at representative early, intermediate, and late times.
}
\label{fig:cutoff}
\end{center}
\end{figure}

\textit{Results}.--- Our main results  are presented in Fig.~\ref{fig:mainresults}(b,c) and Fig.~\ref{fig:cutoff}. Figure~\ref{fig:mainresults}(b) shows the time evolution of the level statistics and the mutual information between the two edge edges for RUCs, while Fig.~\ref{fig:mainresults}(c) shows the same data but for Hamiltonian dynamics (\ref{eq:Isingham}); note the broad similarities between the two sets of data, despite the absence of the strict light cone in the latter. To succinctly characterize the level statistics via a single parameter, we use the so called $r$-ratio, defined as the average over the entanglement spectrum and disorder realizations of $r=\frac{\min(\delta_n , \delta_{n+1})}{\max(\delta_n , \delta_{n+1})}$, where $\delta_n=E_n-E_{n-1}$ is the spacing between consecutive entanglement energy levels~\cite{oganesyan_huse}. The value $r$ is a measure of level repulsion: for Poisson statistics, $r\approx0.39$, whereas for random matrices in the Gaussian Unitary Ensemble, $r\approx 0.6$ {[although the Hamiltonian~\eqref{eq:Isingham} is real, the time-evolved state and hence its entanglement Hamiltonian are generically complex, so the unitary ensemble is the appropriate one]}. 
As a proxy for the mutual information between the left and right halves of the outside world, we take the mutual information between two sites just outside the block $A$ on the left/right, which we denote by ${I}(-l/2,l/2)$.

The level statistics shows a regime of Poisson behavior  after the initial transient, but crosses over to RMT-like behavior at a time that scales linearly with the subsystem size $l$. Note that for the RUC, with a strict light cone, the mutual information between the boundary spins remains exactly zero until this time, when it begins to grow. 
In both Fig.~\ref{fig:mainresults}(b) and (c), we have cut off  the high-energy part of the entanglement spectrum when computing $r$ and include only eigenvalues $E<10$. A more fine-grained picture of entanglement level statistics is provided by studying the time  evolution of the energy-resolved $r$-ratio, takings its average within some small energy window $[E,E+\Delta E]$. 
For the non-integrable Ising model (Fig.~\ref{fig:cutoff}), high entanglement energies exhibit RMT behavior at relatively short times compared to low ones, with the 
`edge' between the two moving roughly linearly in time~\cite{suppmat}. Representative line-cuts of the data at $t=3.4, 4.0, 4.6$ are shown in the bottom panel of Fig.~\ref{fig:cutoff}.  
The discrete time evolution of RUCs makes their entanglement spectral crossover abrupt and challenging to capture on the relatively modest system sizes considered here, though it is qualitatively similar.
Finally, we note that the entanglement entropy~\cite{suppmat} remains far from its thermal volume law value even after the entire entanglement spectrum shows RMT behavior, consistent with our three-stage scenario for thermalization. 

\begin{figure}[t]
\begin{center}
\includegraphics[width = \columnwidth]{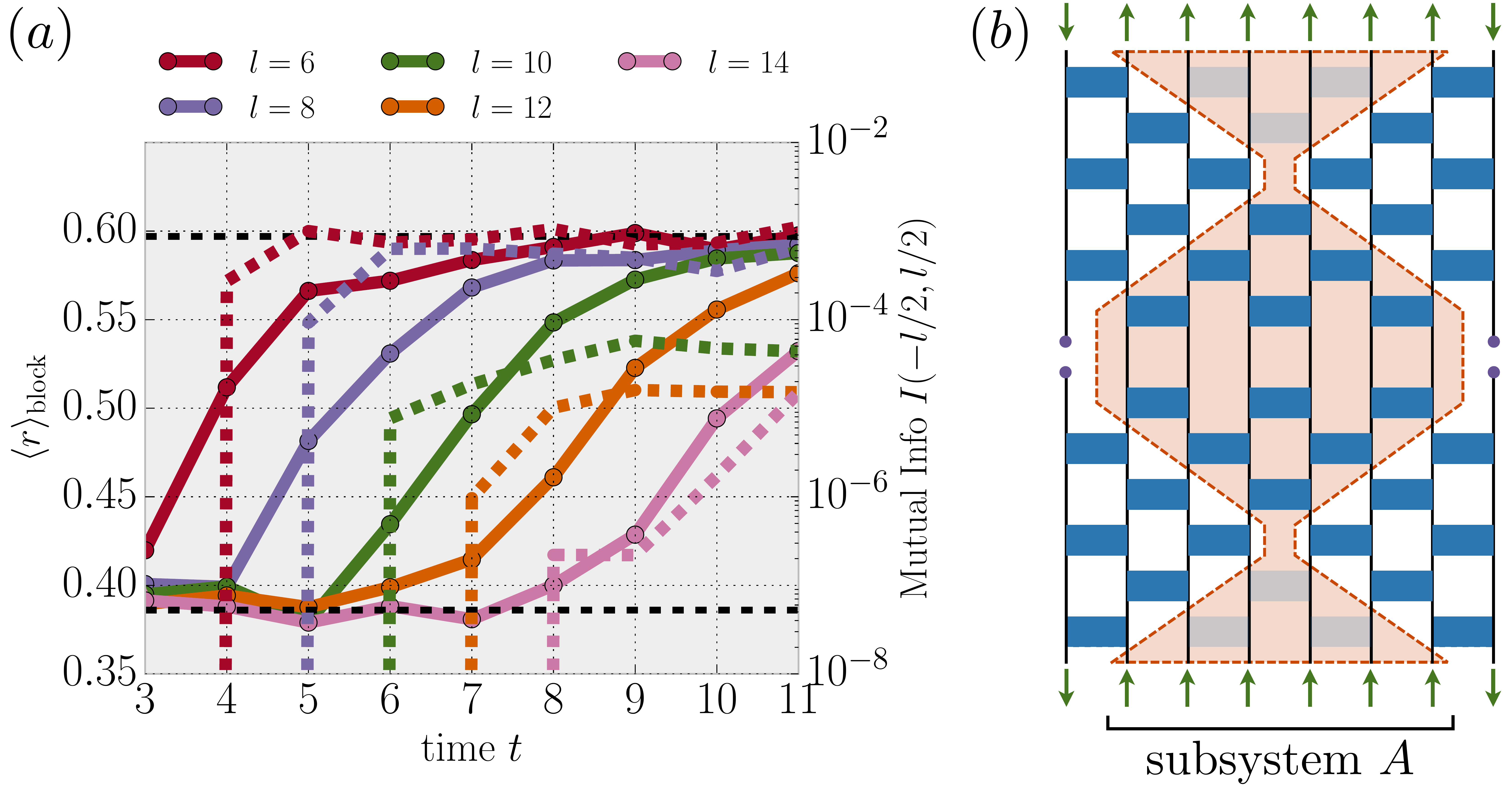}
\caption{(a) Evolution of entanglement level statistics (for entanglement energies $E < 10$) and boundary-spin mutual information in the charge-conserving random circuit, starting in an domain initial  state where the $l$ sites in the subsystem $A$ are occupied (up) and all other sites are empty (down). The time scale of the transition towards random matrix statistics remains linear in $l$, despite completely diffusive charge transport. However, the time scale increases by a factor of two compared to e.g. a Neel initial state, due to the fact that the initial state in the block is an eigenstate of the time evolution. (b) This can be understood by an argument similar to that for the non-conserving case by noting that there is an additional cancellation from the fact that the state inside the domain is an eigenstate of time evolution, so that the lightly shaded gates are pure phase gates that produce no entanglement.}
\label{fig:domain}
\end{center}
\end{figure}

\textit{Conservation Laws.}---So far, we have focused either on random unitary dynamics with no conservation laws, or on Hamiltonian systems where the only conserved quantity is the energy. While energy diffusion can affect the low-energy entanglement spectrum~\cite{EntanglementDiff}, this only affects significantly a small subset of eigenvalues and we do not expect it to show up in the spectral diagnostics presented here. A natural question to ask is if the evolution of the entanglement spectra for certain classes of initial states {\it can} be constrained by conservation laws. It is known, for instance, that the spreading of operators is sensitive to the conservation law, that force change transport to be diffusive and thereby lead to hydrodynamic long time tails in  operator dynamics~\cite{OTOCDiff1,OTOCDiff2}.
 Accordingly, we simulate a random unitary circuit with a $U(1)$ symmetry, where the relevant charge within a region can be viewed as counting the number of up spins in that region. We consider a domain initial state where all spins within region $A$ are up, and all those outside of it are down. Fig.~\ref{fig:domain}(a) shows the resulting entanglement spectrum dynamics, as well as the boundary-spin mutual information. Observe that the time scale for the transition towards RMT behavior as seen in the $r$-ratio continues to coincide with the onset of non-zero mutual information between the boundary spins, and both remain linear in $l$ despite the diffusive charge dynamics. However, we see that the scale for the Poisson-RMT crossover is now increased by a factor of 2 relative to the non-conserving case, to $t=l/v_{\text{LC}}$. 
 
To understand why this is the case, first observe that for times $0\leq t\leq l/2v_{\text{LC}}$, we may simply use the same sequence of arguments as for the non-conserving case (see Fig.~\ref{fig:mainresults}a). For times $l/2v_{\text{LC}} \leq t\leq l/v_{\text{LC}}$, the disentangled region constructed by the backward lightcone argument does not partition the circuit into disjoint regions and  so for a generic circuit the density matrix does not factorize. However, the local conservation law strongly constrains the dynamics, as each 2-site gates  only acts nontrivially on (i.e., entangles) spins when they are anti-parallel. It follows from this that that the dynamics deep within region $A$ must be essentially trivial at early times, and can only multiply the system by an overall phase. Accordingly, the lightly shaded gates in Fig.~\ref{fig:domain}b do not contribute to the entanglement, and may be ignored. Evidently, this picture allows us to construct a ``forward disentangled region'' where the gates act trivially, drawing lightcones inward from the ends of subsystem $A$ to the point $(x,t)= (l/2, \leq l/2v_{\text{LC}})$. Combining these, we see that the entanglement spectrum factorizes as long as the two disentangled regions intersect, i.e. for all times $t\leq l/v_{\text{LC}}$. Intuitively, this is the time for the lightcone emanating from one edge to reach the opposite edge. For times longer than this, the two regions no longer intersect, and the gates that lie in the waist between them will spoil the factorization and drive RMT behavior of the entanglement spectrum. We note that the timescale needed for the high entanglement energies to develop level repulsion remains linear, despite the fact that the entanglement growth at the edges of the domain is initially sub-ballistic~\cite{EntanglementDiff}. However, we observe that the later stage of the process, namely the approach to RMT level statistics is slower than for the non-conserving case (cf. Fig.~\ref{fig:mainresults}(b); we attribute this to the slow mode associated with charge diffusion in the conserving circuit.

\textit{Discussion}.---Our  results have shed light on various aspects of how the density matrix of a generic chaotic quantum system thermalizes. 
Our main window into this question was the spectral statistics of the entanglement Hamiltonian of a block with two ends: this allowed us to explore the transmission of quantum information across the block. We found a correspondence between the onset of level repulsion in the entanglement spectrum of the block and that of mutual information between the two 
regions flanking the block. A striking numerical observation was that the small high entanglement energies (i.e., the small eigenvalues of the RDM) are the first to be 
coupled across the block. Thus the energy-dependence of the entanglement level statistics offers a promising if unusual diagnostic for the shape of the operator front.

Our results suggest many avenues for future study; a particularly direct one is the extension to higher dimensions $d$. In $d > 1$ a generic cut no longer disconnects the complement of a subsystem. However, it is clear that the entanglement spectrum of an infinite strip will be Poisson as discussed here; it therefore seems plausible that regions of sufficiently high aspect ratio will exhibit Poissonian entanglement level statistics. Whether an entanglement delocalization transition occurs for aspect ratios of order unity is unclear, however, and we defer this question to future work. Other natural extensions involve the temperature-dependence of the onset of random-matrix level statistics, as well as the evolution of entanglement level statistics for a many-body localized system.

A direct implication of our results is that the bulk of the entanglement spectrum behaves thermally well before entanglement itself has saturated. This suggests that the late-time dynamics of entanglement must be related to the low-energy \textit{edge} of the entanglement spectrum, and therefore have to do with its extreme-value statistics, which will be addressed elsewhere.

\noindent{\textit{Acknowledgments.---}}  All the authors acknowledge the hospitality of the Kavli Institute for Theoretical Physics at the University of California, Santa Barbara, which is supported from NSF Grant PHY-1748958, where this work was initiated. SG thanks  P.-Y. Chang, X. Chen, A. Lamacraft, and J. H. Pixley for collaboration on related work. FP and TR thank C. W. von Keyserlingk for collaboration on related projects. SG acknowledges support from NSF Grant No. DMR-1653271. FP and TR acknowledge the support of the DFG Research Unit FOR 1807 through grants no. PO 1370/2- 1, TRR80, and the European Research Council (ERC) under the European Union's Horizon 2020 research and innovation program (grant agreement no. 771537).

\bibliography{global.bib,references.bib}
\bibliographystyle{apsrev4-1}

\onecolumngrid
\appendix

\newpage
\begin{center}
{\large \textbf{Supplementary Material for ``Signatures of information scrambling in the dynamics of the entanglement spectrum''\\}}
\end{center}

\section{Perturbative argument for evolution of energy-resolved entanglement spectrum}\label{app:analytical}

In this appendix we give a more detailed account of the analytical arguments presented in the main text regarding the time evolution of the entanglement spectrum. Using existing results, based on Lieb-Robinson bounds, we argue that for any arbitrarily chosen threshold $\epsilon$ one can find a corresponding velocity $v(\epsilon)$ such that the approximate factorization $\rho_{\bar{A}} = \rho_L \otimes \rho_R + \mathcal{O}(\epsilon)$ holds for times $t \leq l/2v(\epsilon)$. Then we argue perturbatively that the effect of the $\mathcal{O}(\epsilon)$ correction on the level statistics is more significant at higher then at lower entanglement energies (i.e., it is most significant for small eigenvalues of $\rho_A$). This implies that focusing on a narrow range of energies around some value $E$ fixes the required precision $\epsilon(E)$, which in turn sets a velocity scale $v(\epsilon(E)) = v(E)$. At times $t < l/2v(E)$ the level statistics around this energy is well described by that of the factorized density matrix and therefore does not show level repulson, while at times $t > l/2v(E)$ the two edges become connected and the block acquires random matrix statistics, as discussed in the main text. We argue that $v(E)$ should be an increasing function of $E$, starting at $v_B$ and either diverging for large $E$, or approaching the strict light cone velocity in cases when it exists (random circuits, certain Floquet systems).

\begin{figure}[H]
 \centering
  		\includegraphics[width=0.6\textwidth]{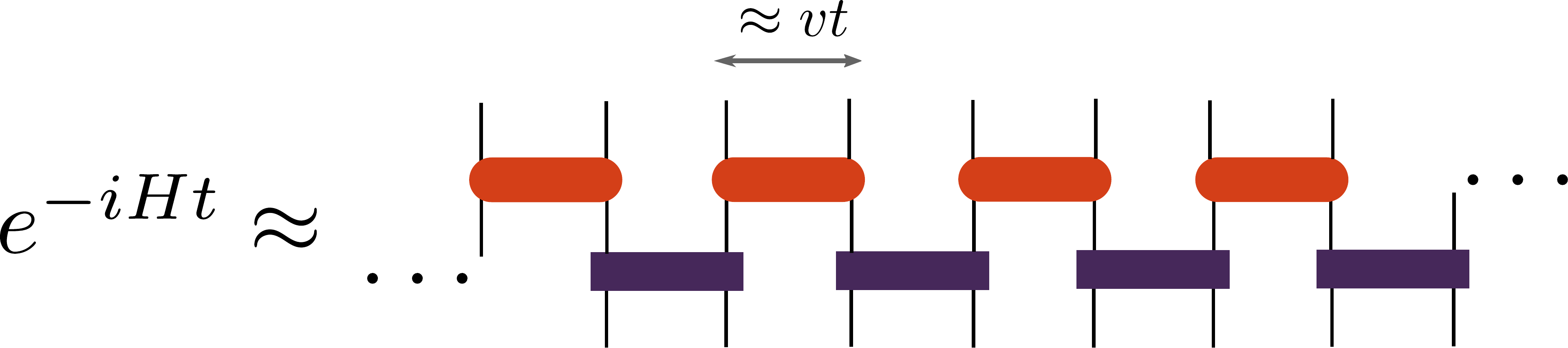}
 \caption{Approximating the time evolution generated by a local Hamiltonian with a two-layer circuit of local unitaries, as described in Ref. \cite{OsborneApprox}.}
 \label{fig:approx}
 \end{figure}

The starting point of our argument is to approximate the time evolution operator $U(t)$ using the results of Ref. \cite{OsborneApprox} as 
\begin{equation}\label{eq:Uapprox}
U = \tilde U + \delta U,
\end{equation}
where $\tilde U$ is a unitary circuit made up by two layers of unitaries, each acting on $vt$ contiguous spins for some constant $v$ (i.e. a block with size linear in $t$) and the the error term has a small operator norm $||\delta U|| = \epsilon$. This approximation is illustrated in Fig.~\ref{fig:approx}. Note that this approximation is done in the spirit of Lieb-Robinson bounds: there is a smallest possible velocity (which we will identify with the Lieb-Robinson velocity) such that the approximation holds, but one can always increase the accuracy by making $v$ larger and thus decreasing the error $\epsilon$. For systems with a \emph{strict} light cone speed $v_\text{LC}$, this approximation becomes exact ($\epsilon = 0$) for $v\geq v_\text{LC}$.

Making use of the above approximation we can write the time evolved state after a quench as
\begin{equation}
\ket{\psi(t)} \equiv U(t) \ket{\psi_0} = \ket{\tilde\psi} + \epsilon \ket{\phi},
\end{equation}
where all the states are normalized to 1 (in principle $\ket{\phi} \equiv \delta U \ket{\psi_0} / \epsilon$ can have some norm $\leq 1$ which we could pull out as a prefactor, but for simplicity we set it to $1$). The corresponding reduced density matrix of a block $A$ of size $l$ is then
\begin{align}
\rho_A = \tilde\rho_A + \epsilon \delta\rho_1 + \epsilon^2 \delta\rho_2; & & \delta \rho_1 \equiv \text{tr}_{\bar A} \left[ \ket{\tilde\psi}\bra{\phi} + h.c.\right]; & & \delta \rho_2 \equiv \text{tr}_{\bar A} \left[ \ket{\phi} \bra{\phi} \right],
\end{align}
with $\bar A$ being the complement of $A$. The first-order correction, $\rho_1$, involves the overlap of two essentially independent vectors on a subsystem of $L-l$ sites; we expect its matrix elements to be of size $\mathcal{O}(2^{(L-l)/2})$, going to zero in the thermodynamic limit. It is therefore safe to neglect this term in the following and focus on the second order correction. 

We consider times at which the light-cone for the  approximate unitary $\tilde U$ has not yet penetrated to the middle of the subsystem $A$. Therefore (see Fig. 1a of main text) $\tilde \rho_A = \rho_L \otimes \rho_R$, so its eigenvalues are of the form $\tilde \Lambda_\alpha = \lambda^L_\alpha \lambda^R_\alpha$. The density of states for $\lambda_L$ is given by $N(\lambda_L) \sim 1/\lambda_L$, and likewise for $\lambda_R$~\cite{RUCEntSpectra}. Combining these, we get

\begin{equation}
N(\tilde \Lambda) = N(\lambda_L) N(\tilde \Lambda/\lambda_L) \sim \int d\lambda \frac{1}{\lambda} \, \frac{1}{\tilde \Lambda/\lambda} \sim \frac{1}{\tilde \Lambda}.
\end{equation}
To leading order, the normalization of $N(\tilde \Lambda)$ is set by the number of nonzero eigenvalues of the RDM, so the density of states at $\tilde \Lambda$ is 

\begin{equation}\label{rhodos}
N(\tilde \Lambda) \approx 2^{2vt} / \tilde\Lambda + \delta(\tilde \Lambda) (2^l - 2^{2vt}),
\end{equation}
up to terms polynomial in $t$ that we neglect~\cite{RUCEntSpectra}. We work in the eigenbasis of $\tilde \rho$ and consider the matrix elements of $\rho_2$ between two eigenstates. This takes the form $\langle \alpha | \phi \rangle \langle \phi | \beta \rangle$, where $|\alpha\rangle, |\beta\rangle$ are Schmidt states of $\tilde \rho$. We approximate $|\phi\rangle$ as a random state, so its overlap with any basis state is of magnitude $2^{-l/2}$, and consequently the typical matrix element of $\rho_2$ between two eigenstates of $\tilde \rho$ has magnitude $\epsilon^2/2^{l}$. To see if nearby energy levels hybridize, we compare this typical matrix element to the energy difference between two adjacent eigenstates of the RDM at energy $\tilde \Lambda$. Any zero modes in Eq.~\eqref{rhodos} will hybridize by this criterion. For the nonzero eigenvalues, hybridization occurs when

\begin{equation}
\tilde \Lambda \leq \epsilon^2 2^{2 v t - l}. 
\end{equation}
Turning now to the entanglement spectrum, we estimate that states with unperturbed energy $\tilde E \equiv -\ln{\tilde \Lambda}$ will develop random-matrix statistics when 

\begin{equation}
\tilde E \geq (l - 2 vt) \log 2 - 2 \log \epsilon.
\end{equation}
To minimize the error while maintaining the separability of $\tilde \rho$ we choose $2 v t = l$. Then the error of the approximation takes the form~\cite{OsborneApprox} $\epsilon \sim e^{(\kappa-\mu v)t} = e^{\kappa t -\mu l/2}$ for some constants $\kappa,\mu$. Plugging in this expression we find that

\begin{equation}
\tilde E \geq \mu l - 2\kappa t.
\end{equation}
This suggests the crossover energy scale from Poisson to RMT statistics should drift linearly downwards in energy as a function of time, consistent with what we see (Fig.~2 of main text).

Note that our entire discussion is independent of the initial state: it therefore provides a lower bound on the times needed for the entanglement spectrum to become RMT-like. We expect that this is the relevant time scale for e.g. a random product state, while the actual time scale can be different for other initial states. We provide an example of this in the main text, where the time scales increase by a factor of $2$ for certain initial states due to conservation laws.

\section{RMT time scale vs. entanglement saturation}

In the main text we claimed that the time needed for the entanglement spectrum (including the lowest energies) to develop level repulsion is parametrically smaller then the time necessary for the block to become fully entangled with the rest of the system. In particular we argued that the the first of these time scales should be controlled by the Lieb-Robinson / butterfly velocity, $v_\text{B}$, that gives the speed at which local operators spread, while the time for the entanglement of the block to saturate is set by the entanglement velocity, $v_\text{E}$, i.e. the rate at which the two sides of a bi-partition become entangled. It is expected on general grounds that the inequality $v_\text{E} < v_\text{B}$ holds~\cite{Mezei16,JonayNahum}, so there should be a time window where the spectrum has already acquired RMT statistics but the amout of entanglement between the block and its environment still keeps increasing. In this appendix we provide numerical evidence for this, in the case of the tilted field Ising model already considered in the main text.

\begin{figure}[H]
 \centering
  		\includegraphics[width=0.35\textwidth]{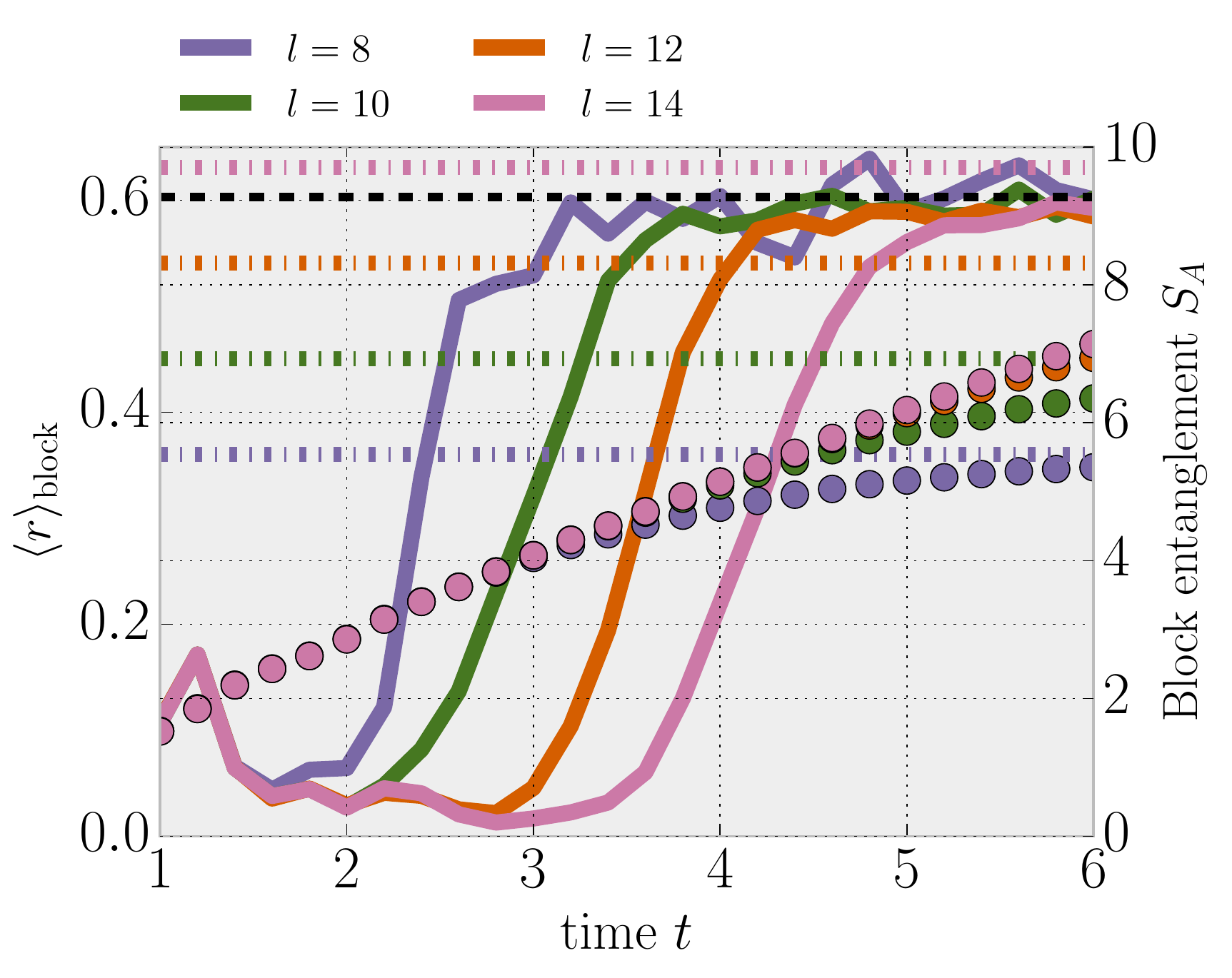}
  		\includegraphics[width=0.35\textwidth]{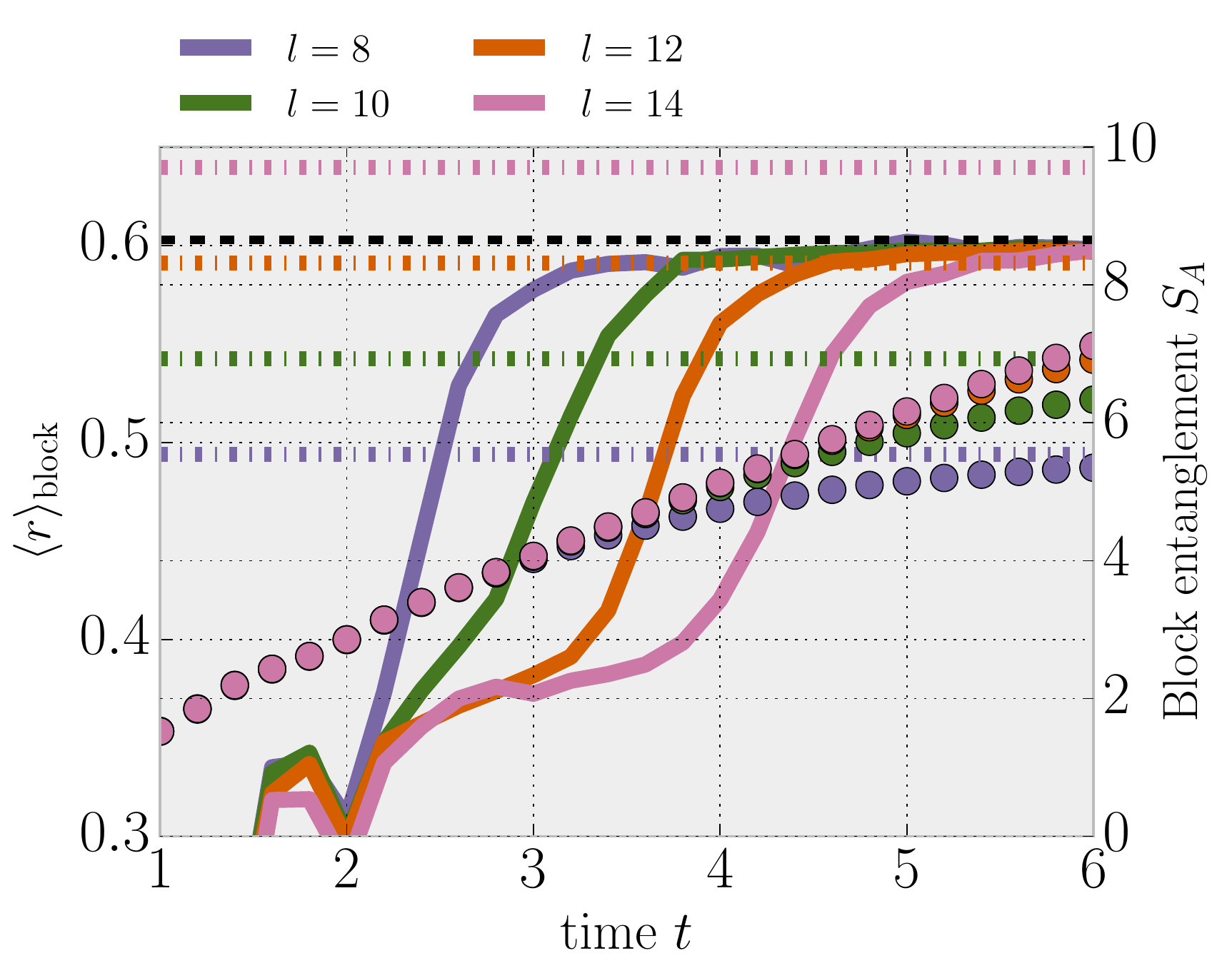}
 \caption{Average $r$-ratio (solid lines) and von Neumann entropy (dots) for blocks of $l$ sites in the tilted field Ising model without disorder (left) and with weak disorder ($W=0.05$, right). The dash-dotted horizontal lines denote the thermal values of the von Neumann entropy $S_A$ at infinite temperature, relevant for the initial Neel state we consider here. At the times when the $r$-ratio saturates to its random matrix value (dashed horizontal line) the total entropy is still far from this thermal value and keeps increasing up to a parametrically longer time scale.}
 \label{fig:IsingEntSat}
 \end{figure}

We take the the Hamiltonian as defined in the main text in and around Eq.~\eqref{eq:Isingham} and simulate its dynamics both in the clean ($W=0$) and the weakly disordered ($W=0.05$) case. We compute the average $r$-ratio of the entanglement spectrum, taking only eigenvalues in the low energy part of the spectrum ($E < 10$), and compare their behavior with the von Neumann entropy of the block $S_A = -\text{tr}(\rho_A \ln{\rho_A})$. As expected, we find that at the times when the $r$-ratio saturates, $S_A$ is still far from its thermal value, as shown in Fig.~\ref{fig:IsingEntSat}. Note that the in the clean, $W=0$, case the $r$-ratio remains close to zero at short times, rather than approaching its Poisson value. This is due to the fact that in this case the Schmidt values at the two edges of the block are identical, resulting in exact degeneracies. 

\section{Numerical results on Floquet model}

To complement the data presented in the main text for random circuits and the tilted field Ising model, here we present some further numerical results on the periodically driven version of the latter, the so-called kicked Ising chain. It is defined through the Floquet unitary that desribes time evolution during a single driving period, which reads 
\begin{equation}\label{eq:floquet_def}
U = e^{-\frac{T}{2}\sum_i h^z_i \sigma^z_i} e^{-\frac{T}{2} \sum_i J_i \sigma^x_i\sigma^x_{i+1} + h^x_i \sigma^x_i}.
\end{equation}
The dynamics generated by repeated application of this unitary can be thought of as being half-way between the aforementioned two models. On the one hand it is simply a (periodically) time-dependent version of the Ising chain described in Eq. (2) of the main text, and has no randomness in the time direction. On the other hand it has no conserved quantities and has a strict light cone velocity of 1 site per Floquet period, which makes it similar to the random circuit model (in fact it can be represented exactly as a circuit with the same geometry). The data presented here emphasizes the universality of our result, which apply to any spatially local time evolution in 1D.

We fix $J_i = 1$ and $T = 1.6$ and choose the on-site fields according to a box distribution of width $W$. We fix the average longitudinal field to be $\overline{h^x} = 0.809$ and consider different values of the average transverse field $\overline{h^z}$. In the clean case it is known that changing the transverse field can be used to tune the butterfly velocity~\cite{RvK17}, between $0$ at $h^z = 0$ and $v_\text{B} \approx v_\text{LC} = 1$ when $h^z \approx 0.9$, and we expect similar dependence on the average transverse field in the weakly disordered case as well. This allows us to explore how the time scales relevant for the block entanglement spectrum change with the butterfly speed and confirm that increasing the latter reduces the time needed to reach random matrix level statistics.

\begin{figure}[H]
 \centering
  		\includegraphics[width=0.35\textwidth]{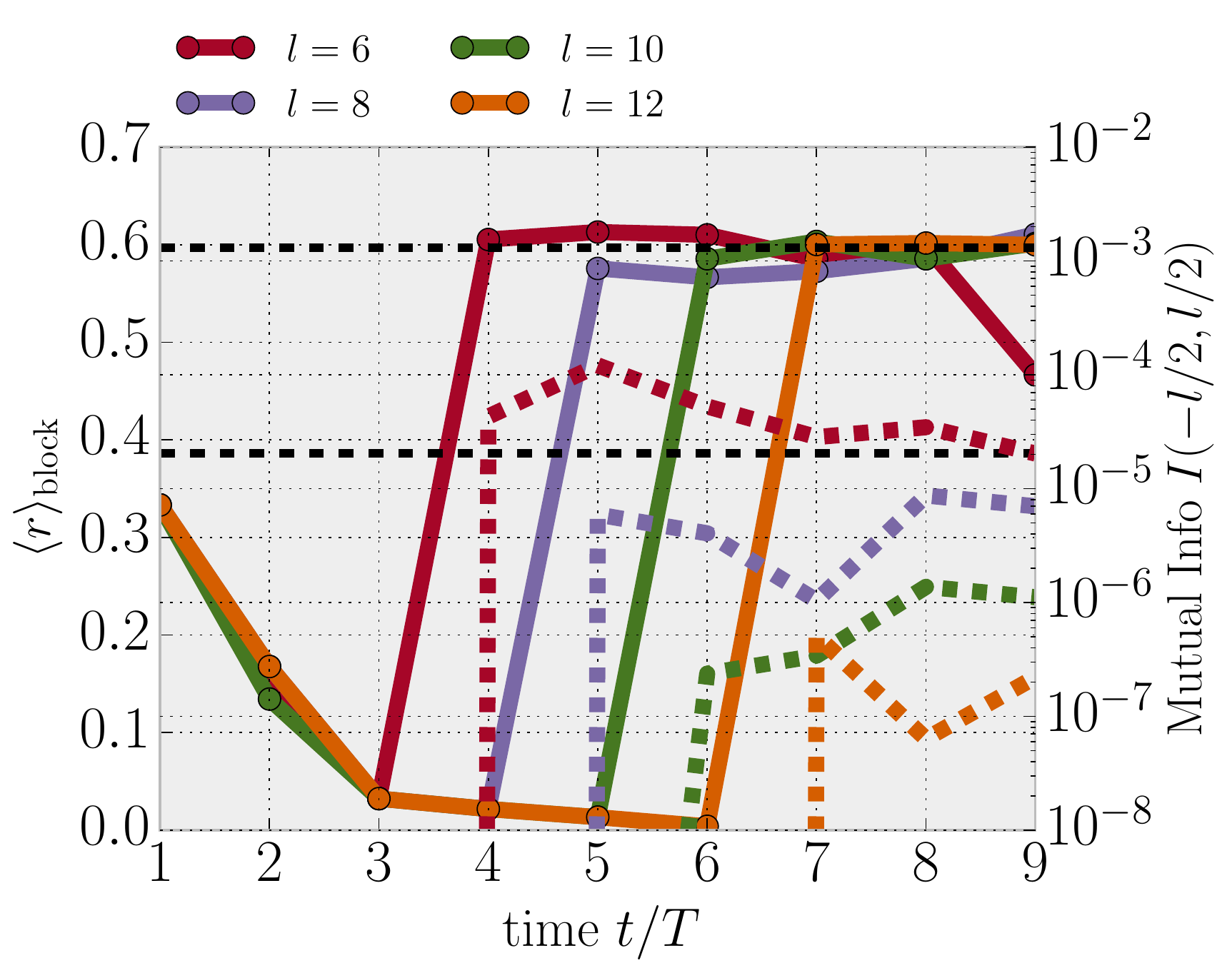}
  		\includegraphics[width=0.35\textwidth]{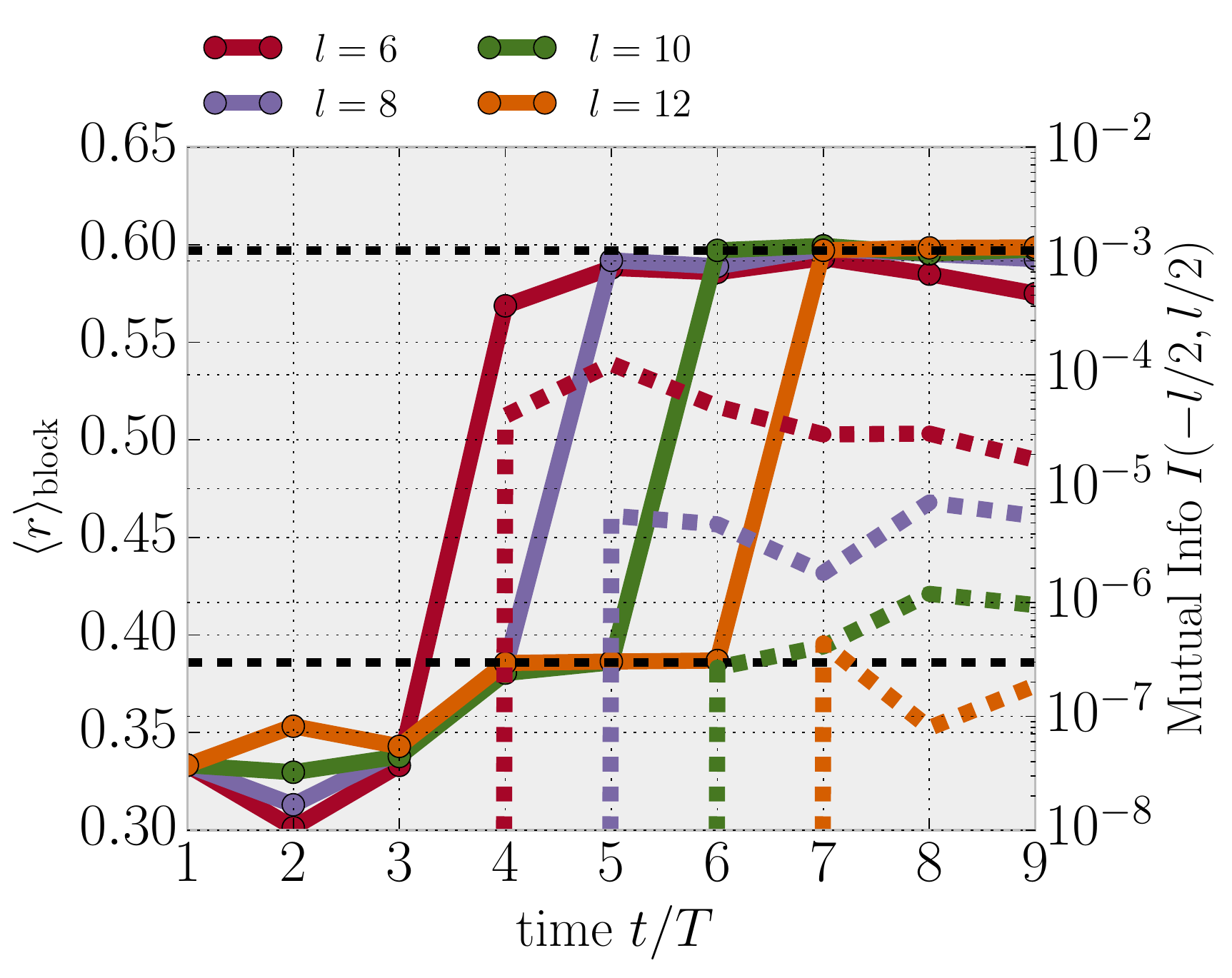}
 \caption{Average $r$-ratio for different block sizes $l$ (solid lines) and mutual information between the two neighboring spins on the left/right side of the block (dashed lines) for the kicked Ising chain defined in Eq.~\eqref{eq:floquet_def} with no disorder ($W=0$, left) and weak disorder $(W=0.05$, right). The average longitudinal field is $\overline{h^z} = 0.7$ in both cases. In calculationg the $r$-ratio, only eigenvalues of the rediced density matrix with magnitude larger then $10^{-10}$ are kept. At this cutoff we observe a sharp transition to random matrix statistics when the strict light cone reaches the middle of the block, at $t=l/2$.}
 \label{fig:floquet1}
 \end{figure}

In Fig.~\ref{fig:floquet1} we show results both for the clean ($W=0$) and weakly disordered ($W=0.05$) chains, comparing level statistics and mutual information as we did for the other models in the main text. By applying a weak cutoff (keeping eigenvalues of $\rho_A$ with magnitude $\Lambda > 10^{-10}$) we observe a sharp transition in the level statistics at times $t=l/2$ when the sharp light cone reaches the middle of the block, similarly to the random circuit case. Before this time the disordered model exhibits Poisson level statistics, also in agreement with our random circuit results. In the clean case, on the other hand, the average $r$-ratio remains close to zero as long as the two edges are uncoupled. This is due to the fact that in this case the entanglement spectra at the two edges of the block are identical, leading to exact degeneracies in the block spectrum, as explained in the main text.

\begin{figure}[H]
 \centering
  		\includegraphics[width=0.3\textwidth]{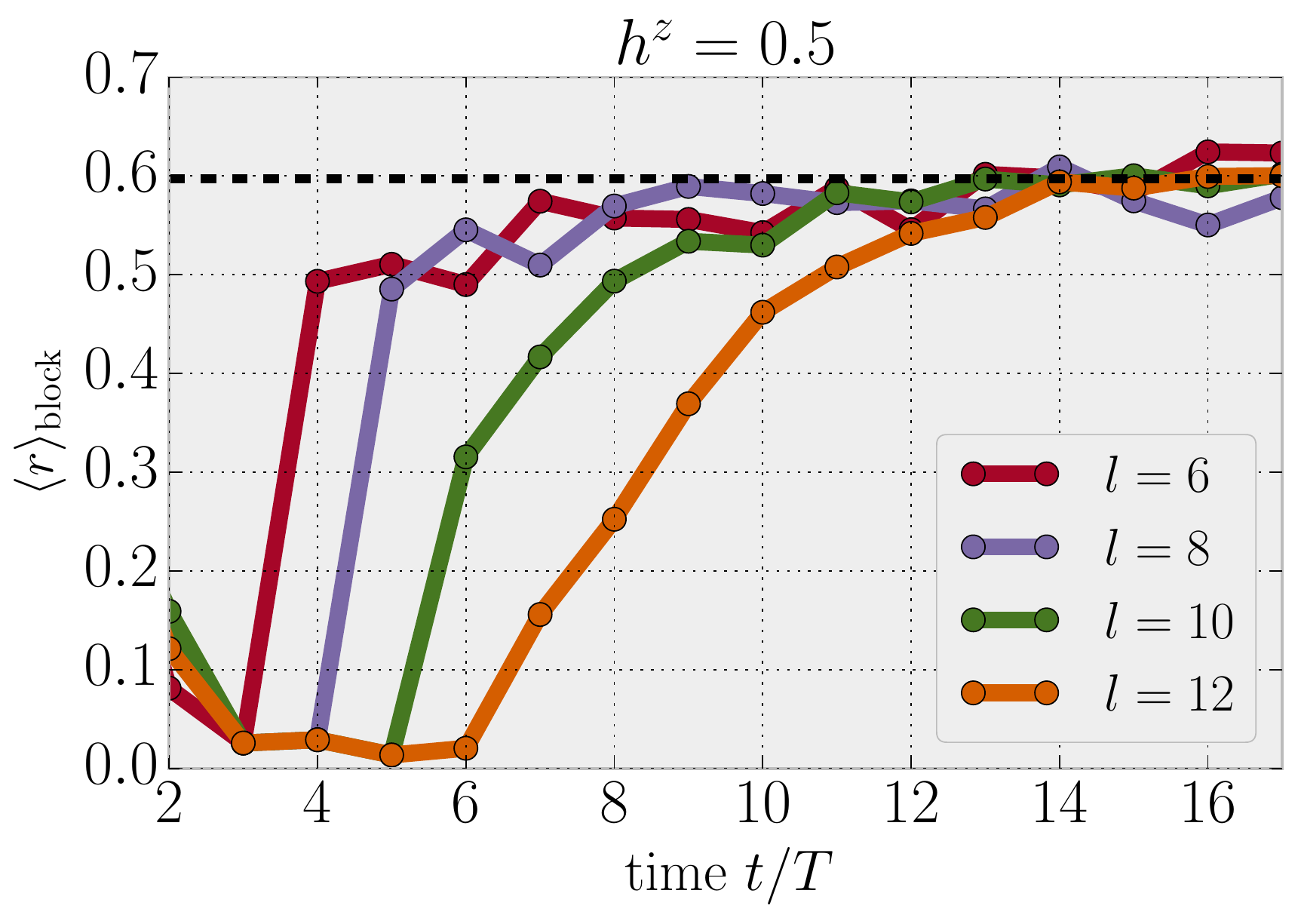}
  		\includegraphics[width=0.3\textwidth]{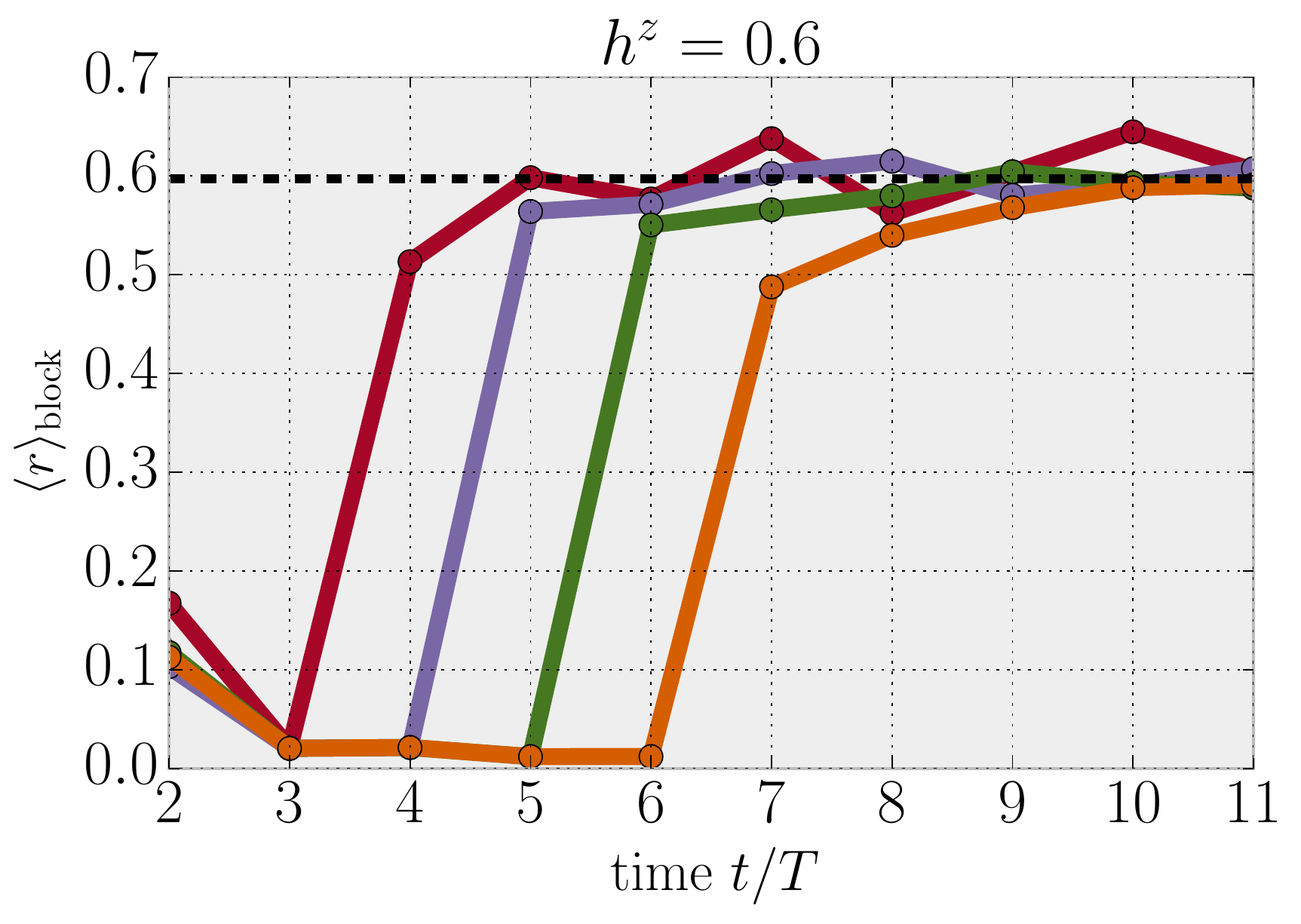}
  		\includegraphics[width=0.3\textwidth]{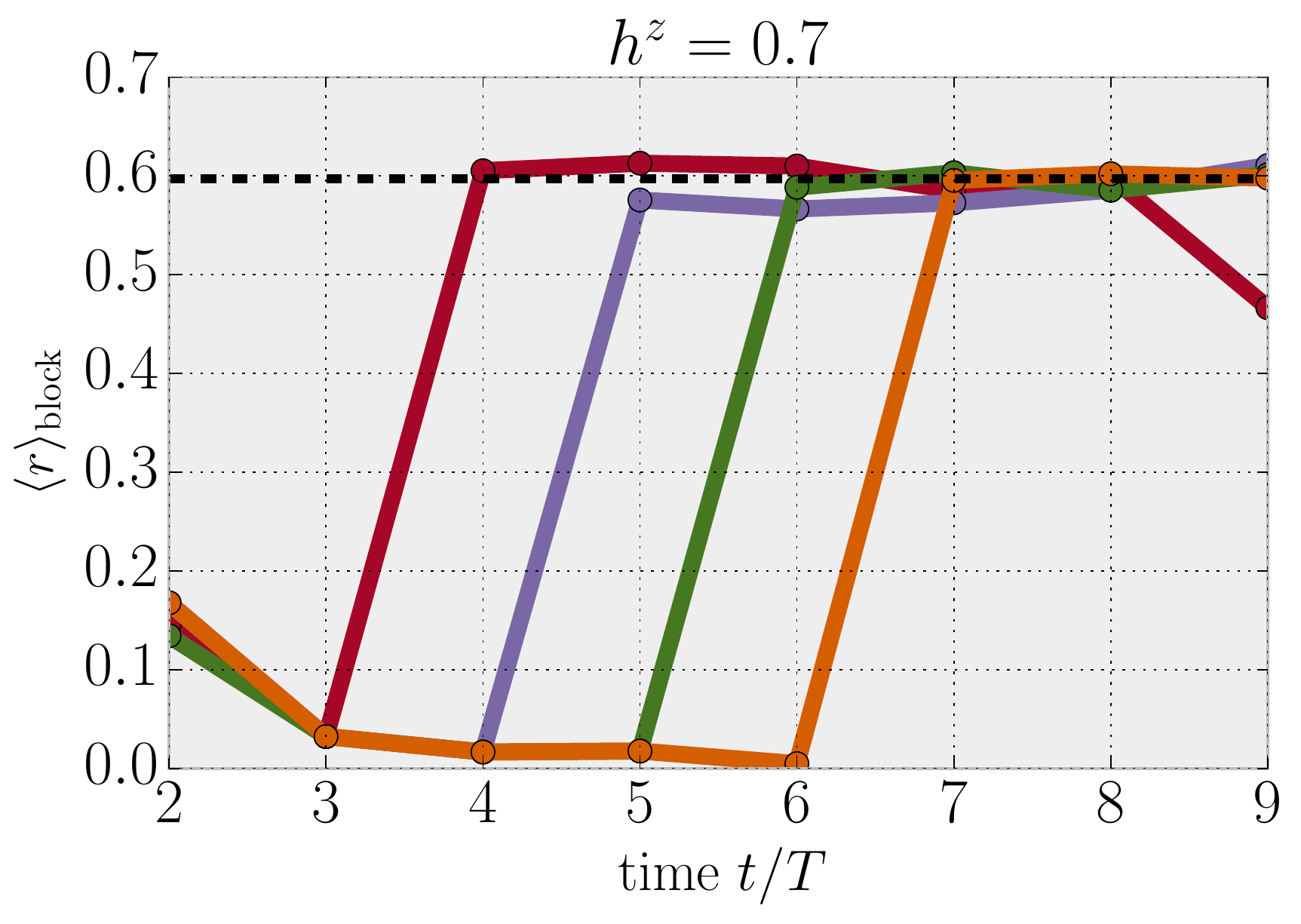}
 \caption{Average $r$-ratio for the clean kicked Ising chain ($W=0$), keeping RDM eigenvalues $>10^{-5}$, for different transverse fields $h^z = 0.5, 0.6, 0.7$ (left to right). While the $r$-ratio starts growing when the strict light cone crosses half the block, the transition is much slower for smaller values of $h^z$, which we attribute to the decrease in the butterfly velocity.}
 \label{fig:floquet2}
 \end{figure}

In Fig.~\ref{fig:floquet2} we show results with a stronger cutoff, keeping only RDM eigenvalues with $\Lambda > 10^{-5}$, for different values of the transverse field $h^z$. We find that, while the transition in the average $r$-ratio always starts at the same time, set by the light cone velocity, it becomes less and less sharp at smaller transverse fields, and the time it takes for $\langle r\rangle$ to reach the random matrix value increases. This can be interpreted by noting that time needed for the spectrum to become fully random matrix-like even at low energies should be controlled by the butterfly / Lieb-Robinson velocity, as detailed in App.~\ref{app:analytical}, which becomes smaller when $h^z$ is decreased as observed previously in Ref.~\cite{RvK17}.

\end{document}